%% file: pressor_response_main.tex
\documentclass{amia}
\usepackage{graphicx}
\usepackage[labelfont=bf]{caption}
\usepackage[superscript,nomove]{cite}
\usepackage{color}

\usepackage{xcolor}
\usepackage{soul}

\usepackage{amsmath}
\usepackage{graphicx}
\usepackage{color}
\usepackage{subcaption}
\usepackage{caption}
\usepackage{geometry}
\usepackage{changepage}

\begin{document}

\title{Predicting Individual Responses to Vasoactive Medications in Children with Septic Shock}

\author{Nicole Fronda, BS$^{1}$, Jessica Asencio MD$^{2}$,  Cameron Carlin, MS$^{1}$, David Ledbetter, BS$^{1}$, Melissa Aczon, PhD$^{1}$, Randall Wetzel, MD$^{1,2}$, Barry Markovitz, MD, MPH$^2$}

\institutes{
    $^1$Laura P. and Leland K. Whittier Virtual Pediatric Intensive Care Unit \\
    $^2$Department of Anesthesiology and Critical Care Medicine  \\
    Children's Hospital Los Angeles, California, United States \\
}

\maketitle

\section*{ABSTRACT}
\input{pressor_response_s0_abstract}

\pagebreak
\section*{INTRODUCTION}
\input{pressor_response_s1_intro}

\section*{MATERIALS AND METHODS}
\input{pressor_response_s2_methods}

\section*{RESULTS}
\input{pressor_response_s3_results}

\section*{DISCUSSION}
\input{pressor_response_s4_discussion}

\section*{CONCLUSION}
\input{pressor_response_s5_conclusion}

\newpage
\bibliographystyle{vancouver}
\bibliography{pressor_response_bibliog}

\newpage
\section*{APPENDIX}
\input{pressor_response_s6_appendix}
\end{document}

%% file: pressor_response_s0_abstract.tex
{\textbf{Objective:}} Predict individual septic children's personalized physiologic responses to vasoactive titrations by training a Recurrent Neural Network (RNN) using EMR data.

{\textbf{Materials and Methods:}} This study retrospectively analyzed EMR of patients admitted to a pediatric ICU from 2009 to 2017. Data included charted time series vitals, labs, drugs, and interventions of children with septic shock treated with dopamine, epinephrine, or norepinephrine. A RNN was trained to predict responses in heart rate (HR), systolic blood pressure (SBP), diastolic blood pressure (DBP) and mean arterial pressure (MAP) to 8,640 titrations during 652 septic episodes and evaluated on a holdout set of 3,883 titrations during 254 episodes. A linear regression model using titration data as its sole input was also developed and compared to the RNN model. Evaluation methods included the correlation coefficient between actual physiologic responses and RNN predictions, mean absolute error (MAE), and area under the receiver operating characteristic curve (AUC).

{\textbf{Results:}} The actual physiologic responses displayed significant variability and were more accurately predicted by the RNN model than by titration alone ($r=0.20\ \text{vs } r=0.05,\, p < 0.01$). The RNN showed MAE and AUC improvements over the linear model. The RNN's MAEs associated with dopamine and epinephrine were 1-3\% lower than the linear regression model MAE for HR, SBP, DBP, and MAP. Across all vitals vasoactives, the RNN achieved 1-19\% AUC improvement over the linear model.

{\textbf{Discussion:}} The wide variability of observed responses indicates that many factors other than dose change affect the response. Not surprisingly, a linear model whose sole input is dose change does not adequately capture the physiologic changes associated with vasoactive titrations, and is no better than chance in predicting individual patient responses. The RNN model showed improvement over the linear model in predicting titration response.

{\textbf{Conclusion:}} This initial attempt in pediatric critical care to predict individual physiologic responses to vasoactive dose changes in children with septic shock demonstrated an RNN model showed some improvement over a linear model. While not yet clinically applicable, further development may assist clinical administration of vasoactive medications in children with septic shock.

%% file: pressor_response_s1_intro.tex
Septic shock in children is a deadly disease with an estimated mortality rate of 5-10\% with aggressive resuscitation \cite{ceneviva1998hemodynamic}. Prior to institution of guidelines for aggressive resuscitation in septic shock, mortality was as high as 50\% \cite{kutko2003mortality}. Current recommendations include the early administration of volume, antibiotics, and vasoactive medications to support and improve end organ perfusion \cite{brierley2009clinical}. Early reversal of septic shock through early resuscitation has been shown to reduce mortality by 9-fold compared to patients who did not receive early resuscitation \cite{han2003early}.


Early reversal of septic shock using fluid and vasoactive therapies are recommended by the American College of Critical Care Medicine\cite{ceneviva1998hemodynamic}. Literature supports the idea of treating children with cold shock with an epinephrine infusion, and children with warm shock on an infusion of norepinephrine\cite{brierley2009clinical}. Dopamine can also be considered as a first line agent for hemodynamic stabilization in septic shock. In patients with meningococcal sepsis, lack of vasopressor therapy administration showed increased odds of death (23.7) compared to patients who received vasopressor therapy\cite{ninis2005role}. The importance of early initiation of vasopressor therapy in patients with fluid-refractory shock cannot be understated. 

Despite these recommendations, there is a paucity of literature concerning the hemodynamic response to vasoactives in children with septic shock. Specifically, the effects of these medications on Heart Rate (HR), Systolic Blood Pressure (SBP), Diastolic Blood Pressure (DBP), and Mean Arterial Pressure (MAP) can be difficult to predict in individual patients, making drug dose titration necessary in a clinical setting. Vasoactives such as dopamine, epinephrine, and norepinephrine can have strong and adverse effects, and anticipating a child's response to them is critical. 

Early, small-scale studies of the cardiovascular effects of vasoactive medications involved healthy adult males who had infusions of epinephrine and norepinephrine. Significant findings included increased cardiac output and systolic blood pressure and decreased peripheral vascular resistance as responses to epinephrine infusion \cite{goldenberg1948hemodynamic}. Patients given an infusion of norepinephrine had uniform increases in systolic blood pressure and mean arterial pressure\cite{goldenberg1948hemodynamic}. These studies paved the way for modern pharmacokinetic studies, which have further elucidated the mechanism of action and function of these endogenous vasopressors. 
Increasing doses of epinephrine during CPR in out-of-hospital cardiac arrests resulted in an overall linear dose response curve in which higher doses of epinephrine resulted in higher systolic and diastolic aortic blood pressures\cite{gonzalez1989dose}. A study of adult patients with septic shock reported linear increases in  heart rate, mean arterial blood pressure, and cardiac index (CI) after infusion of epinephrine as a single agent \cite{moran1993epinephrine}. Norepinephrine was found to have minimal effects on CI or stroke volume index (SVI) in adult patients with septic shock; although in combination with dobutamine, it was found to increase mean arterial pressure, CI, and SVI\cite{martin1999effects}. Martin et. al. compared dopamine and norepinephrine for treating hyperdynamic septic shock in adults and found that patients who received epinephrine had significant increases in MAP and systemic vascular resistance index (SVRI); and in patients who did not respond to high doses of dopamine, addition of norepinephrine produced the previously described changes\cite{martin1993norepinephrine}.

Ceneviva et. al. studied septic children with fluid refractory shock and their response to inotropic/vasoactive therapy \cite{ceneviva1998hemodynamic}. These patients were divided into three groups: those with low cardiac index (CI), those with low systemic vascular resistance index, and those with combined cardiac and vascular dysfunction. They were treated with infusions of inotropes, vasopressors, or vasodilators depending on their physiology. In 44 of 50 patients studied, vasopressor or inotrope therapy was altered as a result of hemodynamic variables found by pulmonary artery catheter measurements. Four of the study patients had a complete change in hemodynamic profile which necessitated a change in vasopressor/inotrope support based on CI and SVRI. This study showed significant heterogeneity in pediatric septic shock throughout the illness course and need for constant re-evaluation of the patient’s hemodynamic state. 

The significant heterogeneity of septic children's responses to vasoactive medications suggests that an individual child's response depends on many variables other than dose titration alone. Age, number of fluid boluses, concurrent medications, steroid use, previous medical history, and underlying illness may all affect the response of a septic patient to vasoactive therapy\cite{overgaard2008inotropes}. No studies, of which the authors are aware, have attempted to account for these variables on an individual level, and this is likely due to the complexity of the problem and limited data. 

The ever increasing availability of electronic medical records (EMR) combined with recent advances in machine learning may enable a more comprehensive analysis of patient data and vasoactive response than previously possible.  This motivated us to hypothesize that a Recurrent neural network (RNN) could extract the relationships between an individual child's EMR and that child's response to vasoactive titration. RNNs are a family of neural network architectures that were designed to process sequential data and incorporate information from previous inputs when calculating predictions\cite{graves2012supervised,goodfellow2016deep}. Successful applications of RNNs include language modeling\cite{sutskever2011generating}, speech recognition\cite{graves2013speech} and machine translation\cite{sutskever2014sequence,bahdanau2014neural}. More recently, RNNs have been applied to problems in the healthcare domain because of their suitability for streaming medical data. These include early detection of critical decompensation in children\cite{shah20162}, onset of heart failure\cite{choi2017using}, dynamic prediction of mortality risk\cite{aczon2017dynamic,ho2017dependence}, de-identification of patient notes\cite{dernoncourt2017identification}, and prediction of individual children's physiologic state associated with successful discharge from a pediatric intensive care unit (PICU)\cite{carlin2017predicting}.

\section*{OBJECTIVE}
The main goal of this study was to train a Recurrent Neural Network (RNN) on EMR data to predict individual septic children's physiologic response (HR, SBP, DBP, and MAP) to vasoactive titration. A secondary goal was to train a linear regression model whos sole input is dose change to provide a baseline against which the RNN model can be compared.

%% file: pressor_response_s2_methods.tex
\subsection*{Data}
This was a retrospective study of Pediatric Intensive Care Unit (PICU) patients at Children's Hospital Los Angeles (CHLA), a tertiary care academic center, between 2009 and 2017. The CHLA Institutional Review Board (IRB) reviewed the study protocol and waived the need for IRB approval. The data were extracted from de-identified observational clinical data collected in Electronic Medical Records (EMR, Cerner, Kansas City, Mo.) and included time series of vital signs, labs, drugs, and interventions, as well as demographic information such as gender, race, and ethnicity. Only patients meeting septic shock criteria -- an elevated or depressed white blood cell count for the patient's age\cite{goldstein6international}, hyperthermia ($> 38.5^{o}C$) or hypothermia ($< 36^{o}C$), and antibiotic administration during their hospitalization -- were included. 1,142 episodes (655 children) meeting septic criteria and were treated with dopamine, epinephrine, or norepinephrine infusions comprised the study cohort.

\subsection*{Response Variables}
\begin{figure}[ht]
  \centering 
  \includegraphics[scale=0.35]{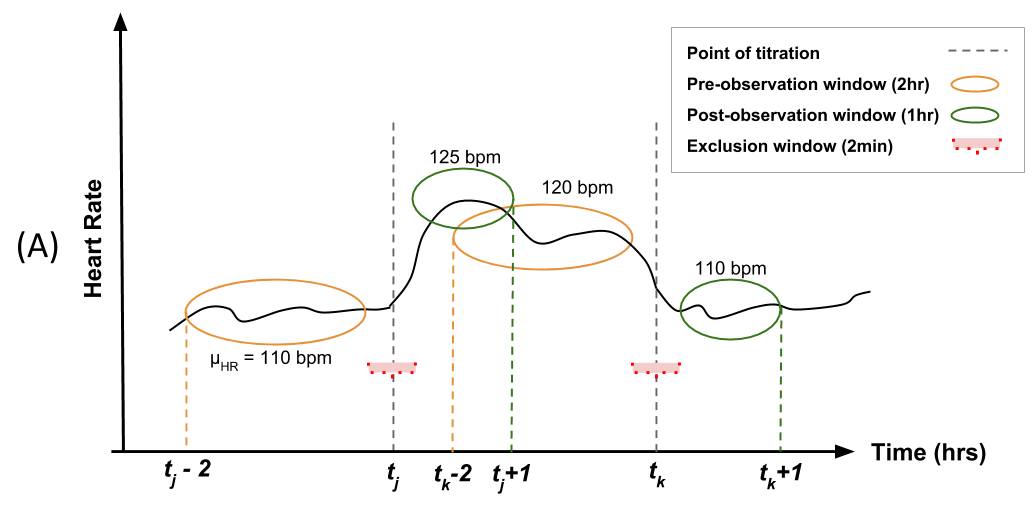}
  \includegraphics[scale=0.55]{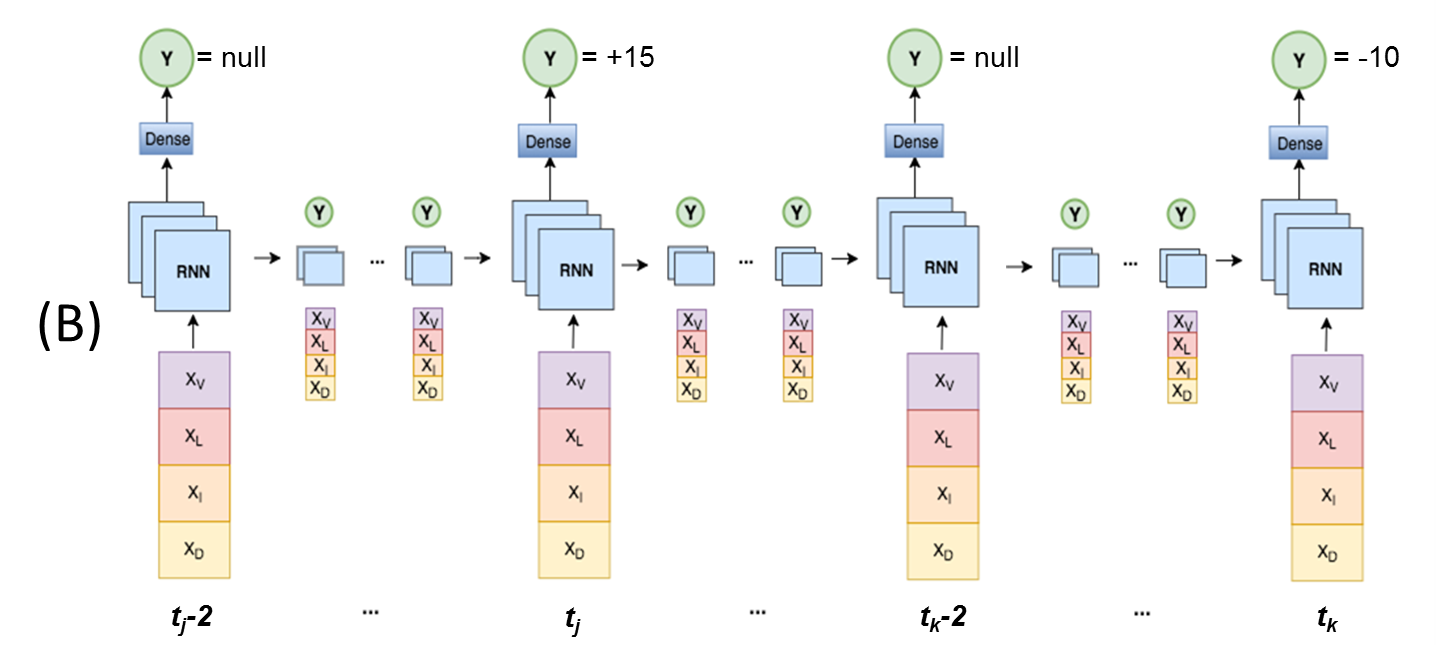}
  \caption{(A) Visualization of how physiologic responses were calculated. In this example, a vasoactive was titrated at times $t_j$ and $t_k$. The average HR in the 2-hour window prior to the 2-minute exclusion window for $t_i$ was 110 bpm. The average HR in the 1-hour window following the 2-minute exclusion window for $t_i$ was 125 bpm. Therefore, the computed true response for the titration at $t_i$ was +15 bpm (125-110).  Similarly, the true response for the titration at $t_k$ was -10 bpm (110-120). (B) High level illustration of a RNN for predicting physiologic response to vasoactive titration.}
  \label{fig:calc_pre_post} 
\end{figure}

HR, SBP, DBP, and MAP were examined prior to and following each \textit{dose titration}, i.e. change in the level of administered dopamine, epinephrine, or norepinephrine. Changes in these four vital signs defined the response we sought to model. These changes were calculated at each titration using the mean measurement one hour after titration (post-observation window) minus the mean measurement two hours before titration (pre-observation window), as shown in Figure \ref{fig:calc_pre_post}A. A two minute exclusion window at the time of administration was used to account for perturbations in actual administration time. 

\subsection*{Linear Regression Model}
Motivated by previous studies \cite{goldenberg1948hemodynamic,gonzalez1989dose,moran1993epinephrine} that modeled physiologic responses to dopamine, epinephrine, and norepinephrine dose changes as linear, we developed linear regression (LR) models of the form $y = mx + b$ as a baseline, where $x$ is the change in vasoactive dose, and $y$ is the change in a vital sign (HR, SBP, DBP, MAP). The training set was used to optimize the model parameters $m$ (slope) and $b$ (intercept). 

\subsection*{RNN Model}
For each vital sign-vasoactive combination, a RNN model using Long Short Term Memory (LSTM) \cite{hochreiter1997long} architecture was developed to predict an individual child's vital sign response to titrations of that vasoactive. At each time point where measurements were recorded, a vector containing patient data at that time point was input to the RNN model. Elements of this vector included vitals, laboratory results, drugs and interventions at that time, as well as demographic information such as gender and race. Preprocessing steps described in previous work\cite{aczon2017dynamic,ho2017dependence} converted the EMR into a temporal sequence of vectors amenable to machine learning. Figure \ref{fig:calc_pre_post}B illustrates the flow of input and output through the RNN. The model makes a prediction at each time point where an input vector exists. The Appendix provides details of the RNN model parameters and training hyperparameters (Table \ref{RNN_param_table}) and a list of input variables (Table \ref{rnn-inputs_5col}).

\subsection*{Model Training}
The error metric used to optimize both LR and RNN model parameters during training was the average of mean absolute error (MAE) of each vital sign aggregated over the training set. Training hyperparameters for the RNN model, including the number of hidden units, optimization algorithm, activation function, and learning rate, were optimized via minimizing MAE on the validation set. Since true responses existed only at times corresponding to points of titration, the errors were aggregated only from the predictions made at these titration points. For example, in Figure \ref{fig:calc_pre_post}, $t_j$ and $t_k$ are the only titration time points. Therefore, only the predictions at these two time points would be included in the error calculation.

\subsection*{Performance Evaluation}
Three metrics computed on the test set assessed model performance. The correlation between the true physiologic response (the measured change in vital sign) and the RNN's predicted change was calculated. This correlation was compared to the correlation between titration dose change and the true physiologic response. The MAEs for each vital sign in each RNN and linear model were also computed. Lastly, the true physiologic responses and predictions were categorized as either increasing or not increasing. The accuracy of these dichotomous predictions was evaluated using the area under the receiver operating characteristic curve (AUC).

\subsection*{Identifying Application and Window Types}
Children may receive various interventions and medications simultaneously during their ICU stay. Consequently, it is difficult to anticipate how an individual child's physiologic state may or may not respond to a certain sequence or combination of treatments, let alone one specific treatment. For each vasoactive of interest, we characterized each titration to aid evaluation of different scenarios in which a child may receive these treatments. This was done by labeling each titration point and each post-observation window as either \textit{independent} or \textit{correlated}.

\begin{figure}[ht]
  \centering 
 \includegraphics[scale=0.35]{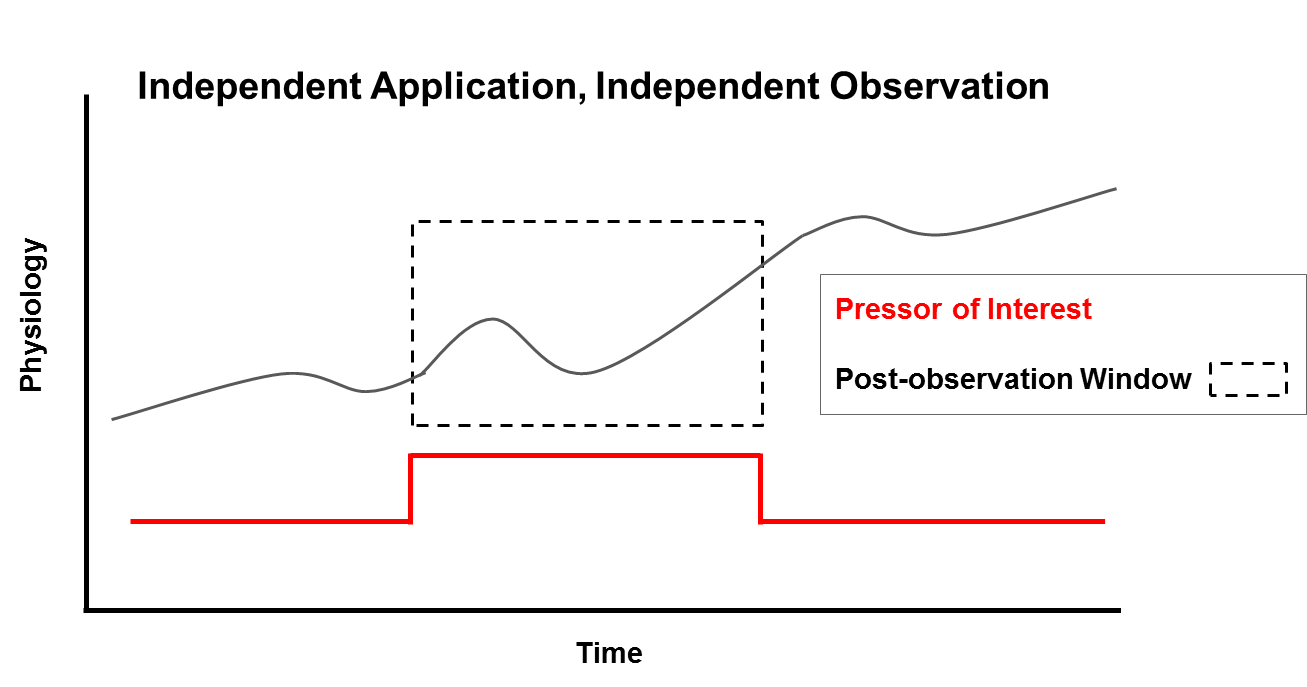}
 \includegraphics[scale=0.35]{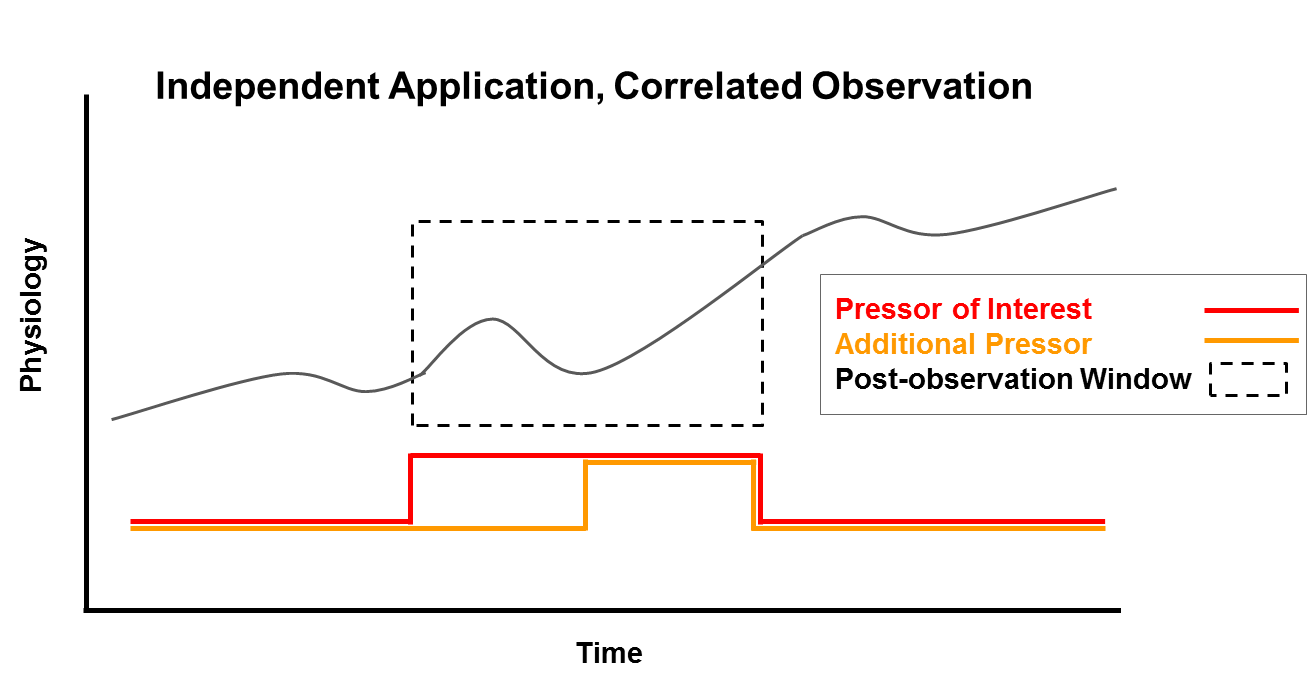}
 \includegraphics[scale=0.35]{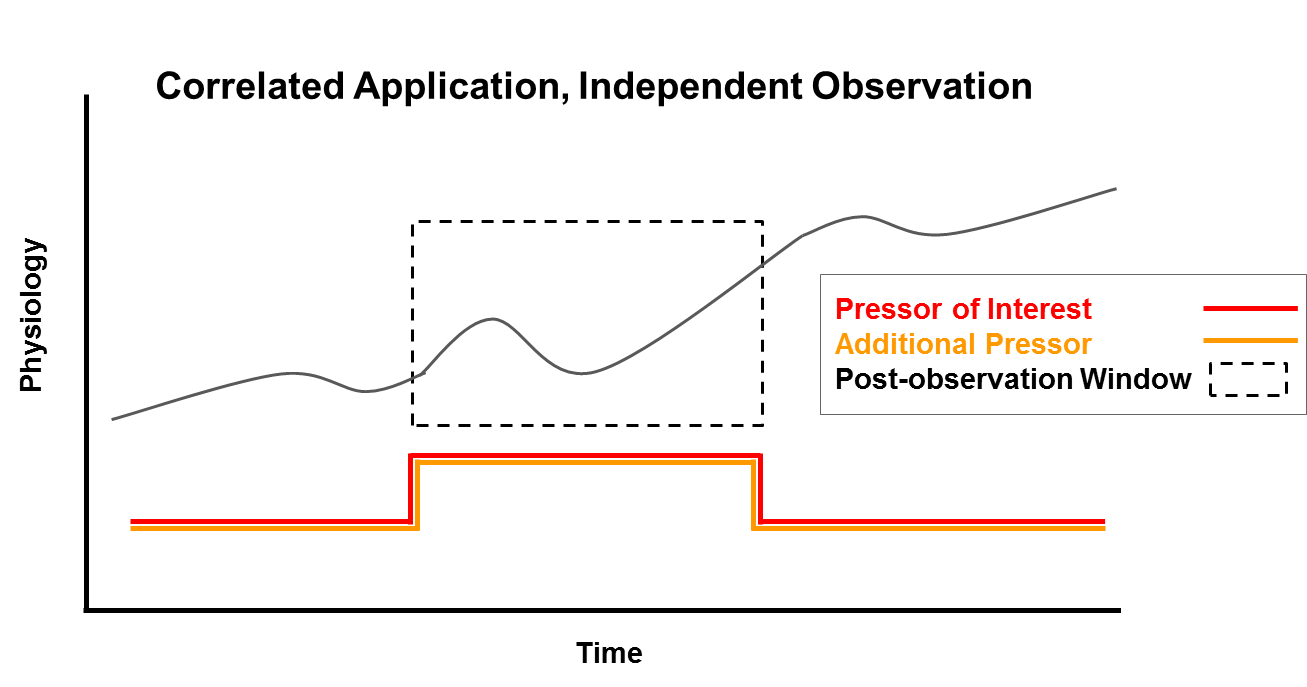}
 \includegraphics[scale=0.35]{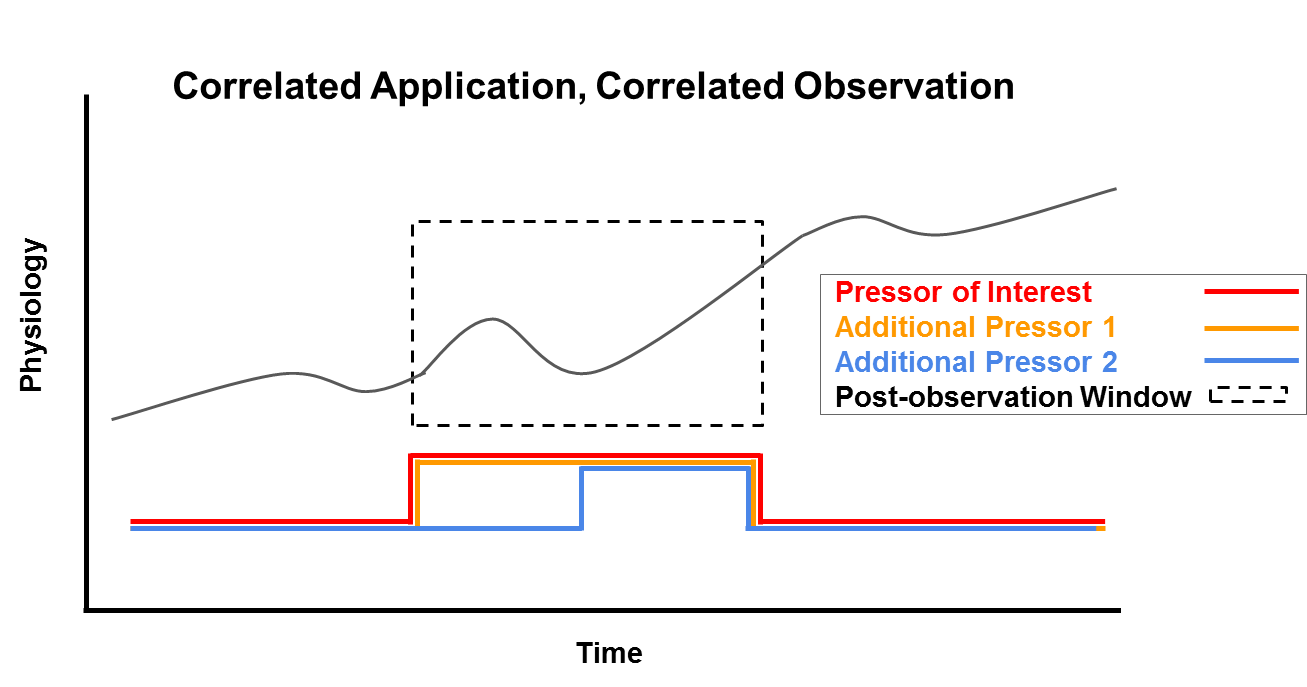}
\caption{Depictions of titration application and observation windows.}
  \label{fig:app_window_types} 
\end{figure}

An \textit{Independent Application} (IA) occurs when exactly one vasoactive of interest -- dopamine, epinephrine, or norepinephrine -- is titrated. A \textit{Correlated Application} (CA) occurs when at least one of these three drugs is titrated at the same time as another vasoactive that may or may not be one of these three. An \textit{Independent Observation} (IO) occurs when the vasoactive of interest is present and no other vasoactives are titrated within the post-observation window. A \textit{Correlated Observation} (CO) occurs when the vasoactive of interest is present and another vasoactive titration occurs within the post-observation window. Figure \ref{fig:app_window_types} illustrates the different scenarios for application and observation.

Characterizing the application and observation type of each titration allows for clearer analysis of model performance. Titrations characterized as an independent application with an independent observation window are the cleanest data for observing the effect of vasoactives. Nevertheless, all application-observation permutations are important to provide a more comprehensive representation of vasoactive use in an ICU setting. Table \ref{table1} provides the counts of application-observation types in the training, validation and test sets for each vasoactive.

\begin{table}[ht]
\centering
\resizebox{\textwidth}{!}{
\begin{tabular}{ll | c | c | c | c | c}
\hline
&  & {Independent} & {Correlated} & Independent & Correlated & IA \& IO \\
Vasoactive & Dataset & Application (IA) &  Application (CA) & Observation (IO) &  Observation (CO) &  \\
\hline
Dopamine & Training & 5,391 & 42 & 3,711 & 1,722 &  3,677\\
 & Validation & 2,069 & 16 & 1,428 & 657  & 1,418 \\
 & Test & 2,422 & 9 & 1,652 & 779 & 1,644\\
\hline
Epinephrine & Training & 2,472 & 20 & 1,462 & 1,030 & 1,450\\
 & Validation & 874 & 6 & 512 & 368 & 509 \\
 & Test & 1,176 & 5 & 694 & 487 &  690\\
\hline
Norepinephrine & Training & 674 & 41 & 383 & 332 & 357\\
 & Validation & 253 & 14 & 142 & 125 & 138 \\
 & Test & 259 & 12 & 147 & 124 & 139\\
\hline
\end{tabular}
}
\caption{Counts of each application and observation type in the training, validation, and test set for each vasoactive.}
\label{table1}
\end{table}

%% file: pressor_response_s3_results.tex
\subsection*{Response Range}
A wide range of responses to vasoactive titrations was observed (Figure \ref{fig:correlation_plots}, left plots). The interquartile range of responses (25\%\,/\,50\%\,/\,75\%) of patients given dopamine was -6/1/4 bpm for HR, -11/2/11 mmHg for SBP, -4/6/8 mmHg for DBP, and -7/2/8 mmHg for MAP per 1 mcg/kg/min increase of dopamine. For 0.1 mcg/kg/min increase of epinephrine, responses were -6/4/9 bpm for HR, -1/12/19 mmHg for SBP, 0/6/13 mmHg for DBP, and 2/8/16 mmHg for MAP. For 0.01 mcg/kg/min increase of norepinephrine, responses were -4/0/4 bpm for HR, -3/3/12 mmHg for SBP, -4/2/7 mmHg for DBP, and -2/2/6 mmHg for MAP. 

\subsection*{Correlation of Response to Dose Change and RNN Prediction}
\begin{figure}
	\centering 
    \includegraphics[scale=0.43]{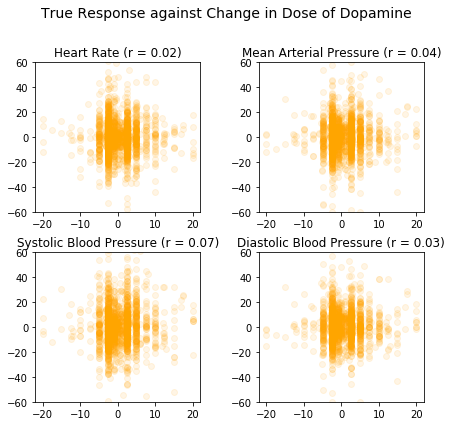}
	\includegraphics[scale=0.43]{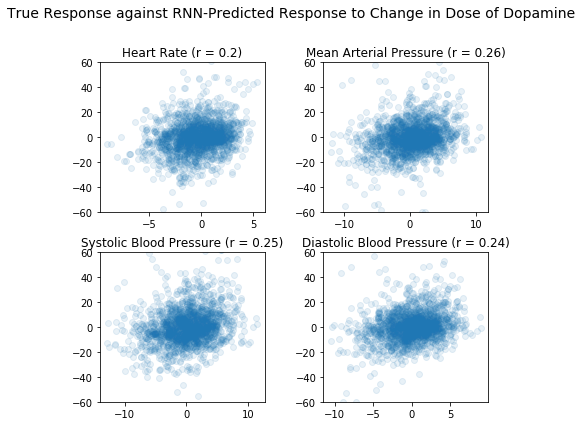}
    \includegraphics[scale=0.43]{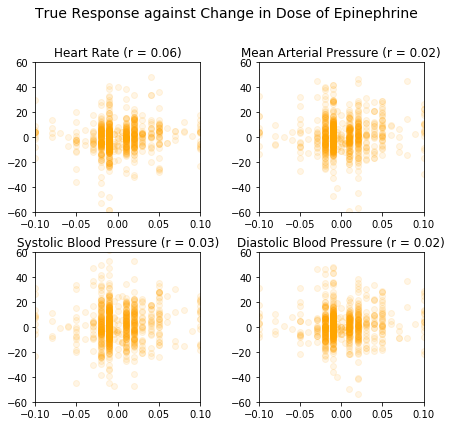}
	\includegraphics[scale=0.43]{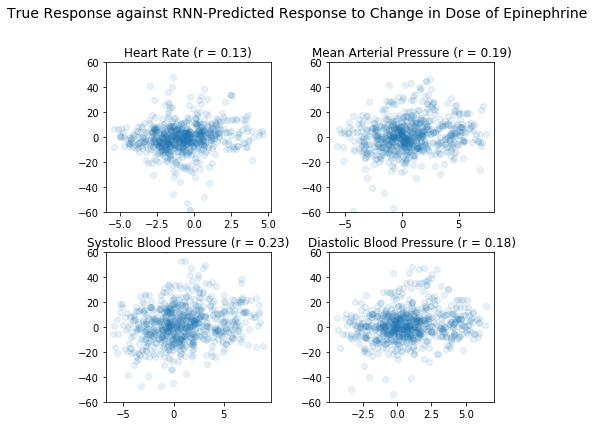}
    \includegraphics[scale=0.43]{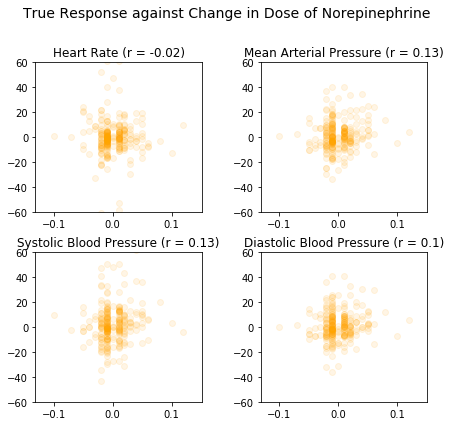}
	\includegraphics[scale=0.43]{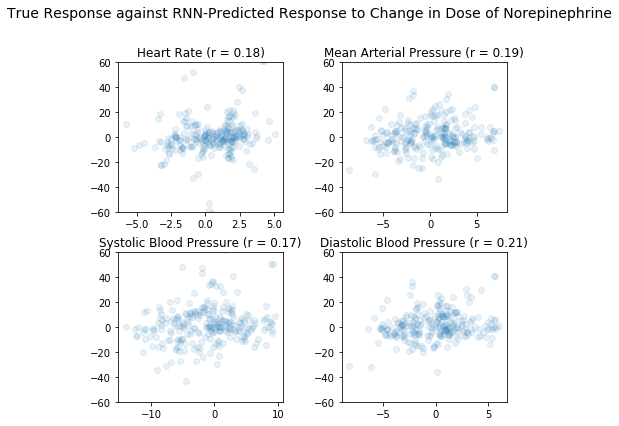}
	\caption{The left plots (yellow) are the vasoactive dose changes in mcg/kg/min (x-axis) and the measured physiologic responses (y-axis). The right plots (blue) are the RNN-predicted responses (x-axis) and the measured responses (y-axis). These plots are shown for each drug (top-bottom: Dopamine, Epinephrine, Norepinephrine). The correlation coefficient (\textit{r}) is included for each subplot.}
	\label{fig:correlation_plots} 
\end{figure}

The correlation coefficient (\textit{r}) between the true responses and dose changes ranged from -0.02 to 0.06, while \textit{r} between the true responses and the RNN's predicted responses ranged from 0.13 to 0.26 (see Figure \ref{fig:correlation_plots}). The difference between these two correlation coefficients was found to be significant ($p < 0.01$). The correlation coefficients between true responses and the titrations characterized as an independent application with independent observation were not significantly different from other titrations (Table \ref{table:corr_coeffs}).

\begin{table}[ht]
\centering
\resizebox{\textwidth}{!}{
\begin{tabular}{l l | rr | rr | rr}
\hline
 &  & \multicolumn{2}{c|} {Over Full Data Set} & \multicolumn{2}{c|} {IA-IO Set} & \multicolumn{2}{c} {Non IA-IO Set} \\\cline{3-8}
 &  & \multicolumn{1}{c}{Response v.} & \multicolumn{1}{c|}{Response v.} & \multicolumn{1}{c}{Response v.} & \multicolumn{1}{c|}{Response v.} & \multicolumn{1}{c}{Response v.} & \multicolumn{1}{c}{Response v.} \\
 &  & \multicolumn{1}{c}{Titration} & \multicolumn{1}{c|}{RNN Pred.} & \multicolumn{1}{c}{Titration} & \multicolumn{1}{c|}{RNN Pred.} & \multicolumn{1}{c}{Titration} & \multicolumn{1}{c}{RNN Pred.} \\
 \hline

Dopamine & Diastolic Blood Pressure & 0.03 & 0.24 & 0.05 & 0.26 & -0.02 & 0.21 \\
 & Heart Rate & 0.02 & 0.20 & 0.01 & 0.18 & 0.01 & 0.26 \\
 & Mean Arterial Pressure & 0.04 & 0.26 & 0.06 & 0.29 & -0.02 & 0.21 \\
 & Systolic Blood Pressure & 0.07 & 0.25 & 0.12 & 0.27 & -0.03 & 0.22 \\
 \hline
Epinephrine & Diastolic Blood Pressure & 0.02 & 0.18 & 0.00 & 0.19 & 0.03 & 0.13 \\
 & Heart Rate & 0.06 & 0.13 & 0.08 & 0.10 & 0.03 & 0.17 \\
 & Mean Arterial Pressure & 0.02 & 0.19 & 0.03 & 0.22 & 0.01 & 0.11 \\
 & Systolic Blood Pressure & 0.03 & 0.23 & 0.04 & 0.22 & 0.01 & 0.20 \\
 \hline
Norepinephrine & Diastolic Blood Pressure & 0.10 & 0.21 & 0.07 & 0.28 & 0.08 & 0.14 \\
 & Heart Rate & -0.02 & 0.18 & -0.05 & 0.19 & 0.03 & 0.16 \\
 & Mean Arterial Pressure & 0.13 & 0.19 & 0.09 & 0.24 & 0.10 & 0.13 \\
 & Systolic Blood Pressure & 0.13 & 0.17 & 0.15 & 0.23 & 0.06 & 0.11 \\
 \hline
\end{tabular}
}
\caption{The correlation coefficient (\textit{r}) for the plots in Figure \ref{fig:correlation_plots}. Correlation coefficients were also calculated separately for titrations characterized as independent application with independent observation (IA-IO) and for titrations not characterized as IA-IO (Non IA-IO).}
\label{table:corr_coeffs}
\end{table}

\subsection*{Evaluation of LR and RNN Predictions Using MAE and AUC}
Parameters of the LR models are in the Appendix (Table \ref{table9}). Table \ref{table-mae} shows the MAEs of the RNN and LR predictions of HR, SBP, DBP and MAP responses to dopamine, epinephrine and norepinephrine titrations in the test set. The RNN's MAEs associated with dopamine and epinephrine were lower by 1-3\% than those of LR across all four vitals. The RNN model had higher MAEs in SBP, DBP and MAP responses to norepinephrine than LR. MAEs aggregated over the subset of titrations characterized as independent application with independent observations were lower than errors aggregated over the whole set.  This was true for both RNN and LR models across all vasoactive-vital combinations.

Table \ref{table-auc} displays test set AUCs when the models were evaluated on the binary task of predicting whether or not vitals increased in response to vasoactive titrations. Across all vitals and vasoactives, the RNN achieved 1-19\% AUC improvement over LR, with the largest improvement coming from HR response to epinephrine. The RNN's AUCs computed over titrations characterized as independent application with independent observation were generally higher than AUCs from the whole set. The only exceptions were dopamine-HR and epinephrine-HR combinations.

\begin{table}[ht]
\centering
\resizebox{\textwidth}{!}{
\begin{tabular}{l | ll | ll | rr | rr}
\hline
MAE & \multicolumn{2}{c|} {HR (bpm)} & \multicolumn{2}{c|}{SBP (mmHg)}  & \multicolumn{2}{c|}{DBP (mmHg)}  & \multicolumn{2}{c} {MAP (mmHg)}  \\
 & \multicolumn{1}{c}{LR} & \multicolumn{1}{c|}{RNN} & \multicolumn{1}{c}{LR} & \multicolumn{1}{c|}{RNN} & \multicolumn{1}{c}{LR} & \multicolumn{1}{c|}{RNN} & \multicolumn{1}{c}{LR} & \multicolumn{1}{c}{RNN} \\
 \hline
Dopamine & 9.34 (9.21) & 9.18 (9.09) & 10.80 (9.99) & 10.51 (9.65) & 8.79 (8.41) & 8.62 (8.17) & 9.38 (8.89) & 9.22 (8.65) \\
Epinephrine & 7.87 (7.79) & 7.77 (7.74) & 11.84 (11.63) & 11.68 (11.42) & 9.13 (8.75) & 9.12 (8.69) & 10.10 (9.63) & 10.02 (9.43) \\
Norepinephrine & 8.57 (7.22) & 8.38 (7.05) & 10.35 (9.38) & 11.15 (9.59) & 8.03 (7.78) & 8.13 (7.54) & 8.47 (8.03) & 8.64 (7.93) \\
\hline
\end{tabular}
}
\caption{Comparison of MAE between the LR and RNN across titration variants. Numbers in parentheses indicate results from only the titrations characterized as independent application with an independent observation.}
\label{table-mae}
\end{table}

\begin{table}[ht]
\centering
\resizebox{\textwidth}{!}{
\begin{tabular}{l | cc | cc | cc | cc}
 \hline
\multicolumn{1}{c|}{AUC} & \multicolumn{2}{c|} {HR (bpm)} & \multicolumn{2}{c|}{SBP (mmHg)}  & \multicolumn{2}{c|}{DBP (mmHg)}  & \multicolumn{2}{c} {MAP (mmHg)}  \\
 & LR & RNN & LR & RNN & LR & RNN & LR & RNN \\
 \hline
Dopamine & 0.528 (0.530) & 0.559 (0.558) & 0.500 (0.502) & 0.564 (0.575) & 0.500 (0.500) & 0.565 (0.575) & 0.499 (0.501) & 0.572 (0.585) \\
Epinephrine & 0.503 (0.503) & 0.597 (0.594) & 0.500 (0.502) & 0.557 (0.563) & 0.498 (0.500) & 0.541 (0.543) & 0.498 (0.500) & 0.532 (0.537) \\
Norepinephrine & 0.528 (0.546) & 0.563 (0.587) & 0.500 (0.500) & 0.506 (0.554) & 0.500 (0.500) & 0.561 (0.577) & 0.500 (0.500) & 0.540 (0.557) \\ \hline
\end{tabular}
}
\caption{Comparison of AUC between the LR and RNN predictions on the binary task across titration variants. Numbers in parenthesis indicate results from only the titrations characterized as independent application and an independent observation.}
\label{table-auc}
\end{table}

%% file: pressor_response_s4_discussion.tex
This is the first attempt to predict an individual critically ill child's response to increasing or decreasing doses of vasoactive medications. The complex environment of an ICU, in which the effects of administered interventions and treatments can vary significantly from what may be observed in a controlled setting, coupled with the complexity of disease processes, pose significant challenges to developing a robust predictive model. 

When defining the response (i.e. vital sign change from pre-titration to post-titration), we deemed a one hour post-observation window sufficient for understanding the vasoactive's effects beyond a child's immediate response to the titration. We also believed the two hour pre-observation window provided more complete information of the child's hemodynamic state before titration to compare with their altered state after titration.

The responses to dopamine, epinephrine and norepinephrine titrations in CHLA's PICU showed significant heterogeneity, consistent with observations from a prior study\cite{ceneviva1998hemodynamic} and the notion that many factors about patient condition affect the response. These results were reflected in the low correlation coefficients between the observed responses and dose changes, regardless of the vasoactive-vital sign combination (Figure \ref{fig:correlation_plots}). Not surprisingly, the linear regression model using only titration information performed poorly in MAE (Table \ref{table-mae}) and AUC (Table \ref{table-auc}), where the latter corresponds to the task of determining positive or negative physiologic response. 

The RNN model, which used all available variables, showed some improvements. Across all vasoactive-vital combinations, the observed responses were more highly correlated to the RNN's predictions than they were to dose changes. The RNN's predictions of all four vital responses to dopamine and epinephrine dose changes showed MAE and AUC improvements over the LR model, while the RNN's predictions of SBP, DBP, and MAP responses to norepinephrine titrations were worse than the regression model's. This is potentially due to the limited norepinephrine data available for training. These results indicate that the RNN model had some ability to extract factors, in addition to vasoactive dose change, that affect response. Both the LR and RNN models performed better on the subset of titrations characterized as independent application with independent observation than on the whole set. This was true for MAEs on all vasoactive-vital combinations, and for AUCs on all but two vasoactive-vital combinations. This affirms our assumption that this subset of titrations would provide a ``cleaner" dataset for understanding the effects of vasoactives, despite the true responses for this subset still showing similar heterogeneity to the full data set.

Future efforts to improve model development include incorporating monitor data and more detailed fluids data. Monitor data captures patient vital sign observations with much higher temporal frequency than the charted data used in this study. Higher fidelity data may provide a more complete picture of a patient's hemodynamic state during any observation window, thus enabling more accurate response computations. In turn, these may allow for better model development. Detailed data for all fluids administered may also improve RNN performance. Fluids are typically administered first before vasoactives to improve blood pressure in septic children, and thus are an important intervention to include in any model.

%% file: pressor_response_s5_conclusion.tex
Septic children's responses to vasoactive titrations displayed significant heterogeneity, which indicates that variables other than dose titration affect the response. We hypothesized that an RNN would be able to process variables and their interactions to more accurately predict an individual child's response than a linear regression using titration data alone. The analysis assessing and comparing the performance of the two models supported this hypothesis. While the RNN's predictions are not yet clinically applicable, the methodologies developed here may provide an initial framework to refine machine learning models for this problem. Further development of these methodologies may assist clinical administration of vasoactive medications in children with septic shock in the future.

%% file: pressor_response_s6_appendix.tex
\begin{table}[ht]
\centering
\begin{tabular}{l l | l | l}
\hline
 & Vitals (units) & Intercept (vital units) & Slope (vital units per dose units) \\
 \hline
Dopamine (mcg) & Diastolic Blood Pressure (mmHg) &  62.010 & 0.002 \\
 & Heart Rate (bpm) & 116.433 & 0.128 \\
 & Mean Arterial Pressure (mmHg) & 75.543 & 0.066 \\
 & Systolic Blood Pressure (mmHg) & 108.157 & 0.242 \\
 \hline
Epinephrine (0.01 mcg) & Diastolic Blood Pressure (mmHg) & 62.240 & 0.056 \\
 & Heart Rate (bpm) & 116.223 & 0.019 \\
 & Mean Arterial Pressure (mmHg) & 75.792 & 0.051 \\
 & Systolic Blood Pressure (mmHg) & 108.542 & 0.060 \\
 \hline
Norepinephrine (0.01 mcg) & Diastolic Blood Pressure (mmHg) & 62.632 & 0.006 \\
 & Heart Rate (bpm) & 116.753 & 0.069 \\
 & Mean Arterial Pressure (mmHg) & 76.202 & 0.068 \\
 & Systolic Blood Pressure (mmHg) & 108.417 & 0.080 \\
\hline
\end{tabular}
\caption{Coefficients for the linear regression models.}
\label{table9}
\end{table}  

\begin{table}[ht]
\centering
\begin{tabular}{l|l}
\hline
Parameter               & Value               \\ \hline
Number of LSTM layers   & 3 \\
Hidden Units in LSTM Layers & 256, 256, 128                 \\
Batch Size              & 32                 \\
Learning Rate           & 0.005             \\
Loss                    & Mean Absolute Error \\
Optimizer               & RMSProp             \\
Dropout                 & 0.2                 \\
Recurrent Dropout       & 0.2                 \\
Regularizer             & L2(0.0001)          \\
Output Activation       & Linear             \\ \hline
\end{tabular}
\caption{RNN Model and Training Parameters}
\label{RNN_param_table}
\end{table}

\begin{table}
\centering
\resizebox{\textwidth}{!}{
\begin{tabular}{lllll}
ABG Base excess & MVBG pH & Dexmedetomidine\_cont & Ondansetron\_inter & Nutrition Level \\ 
ABG FiO2 & Macrocytes & Diazepam\_inter & Oseltamivir\_inter & Oxygenation Index \\ 
ABG HCO3 & Magnesium Level & Digoxin\_inter & Oxacillin\_inter & PaO2 to FiO2 \\ 
ABG O2 sat & Metamyelocytes \% & Diphenhydramine HCl\_inter & Oxcarbazepine\_inter & Patient Mood Level \\ 
ABG PCO2 & Monocytes \% & Dobutamine\_cont & Oxycodone\_inter & Pulse Oximetry \\ 
ABG PO2 & Myelocytes \% & Dopamine\_cont & Pantoprazole\_inter & Quality of Pain Level \\ 
ABG TCO2 & Neutrophils \% & Dornase Alfa\_inter & Penicillin G Sodium\_inter & Respiratory Effort Level \\ 
ABG pH & PT & Enalapril\_inter & Pentobarbital\_inter & Respiratory Rate \\ 
ALT & PTT & Enoxaparin\_inter & Phenobarbital\_inter & Right Pupil Size After Light \\ 
AST & Phosphorus level & Epinephrine\_cont & Phenytoin\_inter & Right Pupil Size Before Light \\ 
Albumin Level & Platelet Count & Epinephrine\_inter & Piperacillin/Tazobactam\_inter & Right Pupillary Response Level \\ 
Alkaline phosphatase & Potassium & Epoetin\_inter & Potassium Chloride\_inter & Sedation Scale Level \\ 
Amylase & Protein Total & Erythromycin\_inter & Potassium Phosphate\_inter & Skin Turgor\_edema \\ 
Anti-Xa Heparin & RBC Blood & Factor VII\_inter & Prednisolone\_inter & Skin Turgor\_turgor \\ 
B-type Natriuretic Peptide & RDW & Famotidine\_inter & Prednisone\_inter & Systolic Blood Pressure \\ 
BUN & Reticulocyte Count & Fentanyl\_cont & Propofol\_cont & Temperature \\ 
Bands \% & Schistocytes & Fentanyl\_inter & Propofol\_inter & Verbal Response Level \\ 
Basophils \% & Sodium & Ferrous Sulfate\_inter & Propranolol HCl\_inter & WAT1 Total \\ 
Bicarbonate Serum & Spherocytes & Filgrastim\_inter & Racemic Epi\_inter & Weight \\ 
Bilirubin Conjugated & T4 Free & Fluconazole\_inter & Ranitidine\_inter & Abdominal X Ray \\ 
Bilirubin Total & TSH & Fluticasone\_inter & Rifampin\_inter & Arterial Line Site \\ 
Bilirubin Unconjugated & Triglycerides & Fosphenytoin\_inter & Risperidone\_inter & CT Abdomen Pelvis \\ 
Blasts \% & VBG Base excess & Furosemide\_cont & Rocuronium\_inter & CT Brain \\ 
C-Reactive Protein & VBG FiO2 & Furosemide\_inter & Sildenafil\_inter & CT Chest \\ 
CBG Base excess & VBG HCO3 & Gabapentin\_inter & Sodium Bicarbonate\_inter & Central Venous Line Site \\ 
CBG FiO2 & VBG O2 sat & Ganciclovir Sodium\_inter & Sodium Chloride\_inter & Chest Tube Site \\ 
CBG HCO3 & VBG PCO2 & Gentamicin\_inter & Sodium Phosphate\_inter & Chest X Ray \\ 
CBG O2 sat & VBG PO2 & Glycopyrrolate\_inter & Spironolactone\_inter & Comfort Response Level \\ 
CBG PCO2 & VBG TCO2 & Heparin\_cont & Sucralfate\_inter & Continuous EEG Present \\ 
CBG PO2 & VBG pH & Heparin\_inter & Tacrolimus\_inter & Diversional Activity\_books \\ 
CBG TCO2 & White Blood Cell Count & Hydrocortisone\_inter & Terbutaline\_cont & Diversional Activity\_music \\ 
CBG pH & Acetaminophen/Codeine\_inter & Hydromorphone\_cont & Tobramycin\_inter & Diversional Activity\_play \\ 
CSF Bands \% & Acetaminophen/Hydrocodone\_inter & Hydromorphone\_inter & Topiramate\_inter & Diversional Activity\_toys \\ 
CSF Glucose & Acetaminophen\_inter & Ibuprofen\_inter & Trimethoprim/Sulfamethoxazole\_inter & Diversional Activity\_tv \\ 
CSF Lymphs \% & Acetazolamide\_inter & Immune Globulin\_inter & Ursodiol\_inter & ECMO Hours \\ 
CSF Protein & Acyclovir\_inter & Insulin\_cont & Valganciclovir\_inter & EPAP \\ 
CSF RBC & Albumin\_inter & Insulin\_inter & Valproic Acid\_inter & FiO2 \\ 
CSF Segs \% & Albuterol\_inter & Ipratropium Bromide\_inter & Vancomycin\_inter & Gastrostomy Tube Location \\ 
CSF WBC & Allopurinol\_inter & Isoniazid\_inter & Vasopressin\_cont & HFOV Amplitude \\ 
Calcium Ionized & Alteplase\_inter & Isradipine\_inter & Vecuronium\_inter & HFOV Frequency \\ 
Calcium Total & Amikacin\_inter & Ketamine\_cont & Vitamin K\_inter & Hemofiltration Therapy Mode \\ 
Chloride & Aminophylline\_cont & Ketamine\_inter & Voriconazole\_inter & IPAP \\ 
Complement C3 Serum & Aminophylline\_inter & Ketorolac\_inter & Age & Inspiratory Time \\ 
Complement C4 Serum & Amlodipine\_inter & Labetalol\_inter & Sex\_F & MRI Brain \\ 
Creatinine & Amoxicillin/clavulanic acid\_inter & Lactobacillus\_inter & Sex\_M & Mean Airway Pressure \\ 
Culture Blood & Amoxicillin\_inter & Lansoprazole\_inter & Abdominal Girth & Mechanical Ventilation Mode \\ 
Culture CSF & Amphotericin B Lipid Complex\_inter & Levalbuterol\_inter & Activity Level & MultiDisciplinaryTeam Present \\ 
Culture Fungus Blood & Ampicillin/Sulbactam\_inter & Levetiracetam\_inter & Bladder pressure & NIV Mode \\ 
Culture Respiratory & Ampicillin\_inter & Levocarnitine\_inter & Capillary Refill Rate & NIV Set Rate \\ 
Culture Urine & Aspirin\_inter & Levofloxacin\_inter & Central Venous Pressure & Nitric Oxide \\ 
Culture Wound & Atropine\_inter & Levothyroxine Sodium\_inter & Cerebral Perfusion Pressure & Nurse Activity Level Completed \\ 
D-dimer & Azathioprine\_inter & Lidocaine\_inter & Diastolic Blood Pressure & O2 Flow Rate \\ 
ESR & Azithromycin\_inter & Linezolid\_inter & EtCO2 & Oxygen Mode Level \\ 
Eosinophils \% & Baclofen\_inter & Lisinopril\_inter & Extremity Temperature Level & Oxygen Therapy \\ 
Ferritin Level & Basiliximab\_inter & Lorazepam\_inter & Eye Response Level & PEEP \\ 
Fibrinogen & Budesonide\_inter & Magnesium Sulfate\_inter & FLACC Pain Activity & Peak Inspiratory Pressure \\ 
GGT & Bumetanide\_inter & Meropenem\_inter & FLACC Pain Consolability & Peritoneal Dialysis Type \\ 
Glucose & Calcium Chloride\_cont & Methadone\_inter & FLACC Pain Cry & Pharmacological Comfort Measures Given \\ 
Haptoglobin & Calcium Chloride\_inter & Methylprednisolone\_inter & FLACC Pain Face & Position Support Given \\ 
Hematocrit & Calcium Gluconate\_inter & Metoclopramide\_inter & FLACC Pain Intensity & Position Tolerance Level \\ 
Hemoglobin & Carbamazepine\_inter & Metronidazole\_inter & FLACC Pain Legs & Pressure Support \\ 
INR & Cefazolin\_inter & Micafungin\_inter & Foley Catheter Volume & Range of Motion Assistance Type \\ 
Influenza Lab & Cefepime\_inter & Midazolam HCl\_cont & Gastrostomy Tube Volume & Sedation Intervention Level \\ 
Lactate & Cefotaxime\_inter & Midazolam HCl\_inter & Glascow Coma Score & Sedation Response Level \\ 
Lactate Dehydrogenase Blood & Cefoxitin\_inter & Milrinone\_cont & Head Circumference & Tidal Volume Delivered \\ 
Lactic Acid Blood & Ceftazidime\_inter & Montelukast Sodium\_inter & Heart Rate & Tidal Volume Expiratory \\ 
Lipase & Ceftriaxone\_inter & Morphine\_cont & Height & Tidal Volume Inspiratory \\ 
Lymphocyte \% & Cephalexin\_inter & Morphine\_inter & Hemofiltration Fluid Output & Tidal Volume Set \\ 
MCH & Chloral Hydrate\_inter & Mycophenolate Mofetl\_inter & Intracranial Pressure & Tracheostomy Tube Size \\ 
MCHC & Chlorothiazide\_inter & Naloxone HCL\_cont & Left Pupil Size After Light & Ventilator Rate \\ 
MCV & Ciprofloxacin HCL\_inter & Naloxone HCL\_inter & Left Pupil Size Before Light & Ventriculostomy Site \\ 
MVBG Base Excess & Cisatracurium\_cont & Nifedipine\_inter & Left Pupillary Response Level & Visitor Mood Level \\ 
MVBG FiO2 & Clindamycin\_inter & Nitrofurantoin\_inter & Level of Consciousness & Visitor Present \\ 
MVBG HCO3 & Clonazepam\_inter & Nitroprusside\_cont & Lip Moisture Level & Volume Tidal \\ 
MVBG O2 Sat & Clonidine HCl\_inter & Norepinephrine\_cont & Mean Arterial Pressure &  \\ 
MVBG PCO2 & Cyclophosphamide\_inter & Nystatin\_inter & Motor Response Level &  \\ 
MVBG PO2 & Desmopressin\_inter & Octreotide Acetate\_cont & Nasal Flaring Level &  \\ 
MVBG TCO2 & Dexamethasone\_inter & Olanzapine\_inter & Near-Infrared Spectroscopy SO2 &  \\ 
\end{tabular}
}
\caption{Input variables to the RNN model.}
\label{rnn-inputs_5col}
\end{table}

\begin{table}[ht]
\centering
\begin{tabular}{l  l | rr | rr | rr | rr}
\hline
\multicolumn{2}{c|} {Dopamine} & \multicolumn{2}{c|} {HR (bpm)} & \multicolumn{2}{c|}{SBP (mmHg)}  & \multicolumn{2}{c|}{DBP (mmHg)}  & \multicolumn{2}{c} {MAP (mmHg)}  \\
&  & LR & RNN & LR & RNN & LR & RNN & LR & RNN \\
 \hline
High Dosage(\textgreater{}10 mcg/kg/min) & MAE & 8.702 & 8.620 & 11.244 & 11.317 & 8.172 & 8.121 & 8.588 & 8.570 \\
 & AUC & 0.560 & 0.572 & 0.486 & 0.574 & 0.500 & 0.586 & 0.500 & 0.613 \\
\hline
Low Dosage(\textless{}10 mcg/kg/min) & MAE & 9.318 & 9.188 & 9.732 & 9.319 & 8.455 & 8.183 & 8.955 & 8.664 \\
 & AUC & 0.523 & 0.556 & 0.502 & 0.573 & 0.500 & 0.573 & 0.501 & 0.578 \\
\hline
High Titration(\textgreater{}5 mcg/kg/min) & MAE & 9.913 & 9.706 & 9.674 & 10.945 & 8.822 & 8.751 & 7.582 & 8.057 \\
 & AUC & 0.486 & 0.598 & 0.676 & 0.588 & 0.500 & 0.611 & 0.576 & 0.632 \\
\hline
Low Titration(\textless{}5 mcg/kg/min) & MAE & 9.194 & 9.074 & 9.995 & 9.616 & 8.395 & 8.155 & 8.933 & 8.666 \\
 & AUC & 0.530 & 0.557 & 0.497 & 0.574 & 0.500 & 0.574 & 0.500 & 0.582 \\
\hline
Dose Increase & MAE & 9.537 & 9.216 & 10.619 & 10.41 & 8.311 & 8.215 & 9.286 & 9.162 \\
 & AUC & 0.512 & 0.605 & 0.500 & 0.586 & 0.500 & 0.575 & 0.500 & 0.582 \\
\hline
Dose Decrease & MAE & 9.025 & 9.018 & 9.617 & 9.215 & 8.464 & 8.147 & 8.665 & 8.349 \\
 & AUC & 0.500 & 0.528 & 0.485 & 0.545 & 0.500 & 0.568 & 0.500 & 0.574 \\
\hline
\end{tabular}
\caption{Comparison of MAE and AUC between the LR and RNN across dopamine titrations characterized as an independent application and independent observation, parsed by titration types.}
\label{table5}
\end{table}

\begin{table}
\centering
\begin{tabular}{l  l | rr | rr | rr | rr}
\hline
\multicolumn{2}{c|} {Epinephrine} & \multicolumn{2}{c|} {HR (bpm)} & \multicolumn{2}{c|}{SBP (mmHg)}  & \multicolumn{2}{c|}{DBP (mmHg)}  & \multicolumn{2}{c} {MAP (mmHg)}  \\
&  & LR & RNN & LR & RNN & LR & RNN & LR & RNN \\
 \hline
High Dosage(\textgreater{}0.1 mcg/kg/min) & MAE & 7.644 & 7.462 & 14.471 & 14.243 & 11.752 & 11.613 & 12.476 & 12.206 \\
 & AUC & 0.514 & 0.681 & 0.500 & 0.561 & 0.490 & 0.555 & 0.491 & 0.535 \\
\hline
Low Dosage(\textless{}0.1 mcg/kg/min) & MAE & 7.826 & 7.815 & 10.842 & 10.630 & 7.908 & 7.875 & 8.832 & 8.660 \\
 & AUC & 0.500 & 0.571 & 0.503 & 0.562 & 0.503 & 0.531 & 0.503 & 0.531 \\
\hline
High Titration(\textgreater{}0.1 mcg/kg/min) & MAE & 11.204 & 11.854 & 12.722 & 11.136 & 7.370 & 6.612 & 8.737 & 7.510 \\
 & AUC & 0.466 & 0.273 & 0.571 & 0.643 & 0.488 & 0.643 & 0.512 & 0.667 \\
\hline
Low Titration(\textless{}0.1 mcg/kg/min) & MAE & 7.663 & 7.589 & 11.599 & 11.426 & 8.788 & 8.753 & 9.653 & 9.492 \\
 & AUC & 0.500 & 0.596 & 0.500 & 0.561 & 0.500 & 0.540 & 0.500 & 0.532 \\
\hline
Dose Increase & MAE & 8.022 & 7.967 & 11.610 & 11.449 & 9.132 & 9.067 & 10.580 & 10.296 \\
 & AUC & 0.505 & 0.595 & 0.500 & 0.524 & 0.500 & 0.523 & 0.500 & 0.511 \\
\hline
Dose Decrease & MAE & 7.656 & 7.611 & 11.645 & 11.400 & 8.535 & 8.483 & 9.105 & 8.961 \\
 & AUC & 0.500 & 0.575 & 0.503 & 0.562 & 0.500 & 0.544 & 0.500 & 0.539 \\
\hline
\end{tabular}
\caption{Comparison of MAE and AUC between the LR and RNN across epinephrine titrations characterized as an independent application and independent observation, partitioned by titration types.
}
\label{table6}
\end{table}

\begin{table}
\centering
\begin{tabular}{l  l | rr | rr | rr | rr}
\hline
\multicolumn{2}{c|} {Norepinephrine} & \multicolumn{2}{c|} {HR (bpm)} & \multicolumn{2}{c|}{SBP (mmHg)}  & \multicolumn{2}{c|}{DBP (mmHg)}  & \multicolumn{2}{c} {MAP (mmHg)}  \\
&  & LR & RNN & LR & RNN & LR & RNN & LR & RNN \\
 \hline
High Dosage (\textgreater{}0.05 mcg/kg/min) & MAE & 5.405 & 5.170 & 8.704 & 9.138 & 8.200 & 7.665 & 7.803 & 7.533 \\
 & AUC & 0.625 & 0.583 & 0.500 & 0.509 & 0.500 & 0.611 & 0.500 & 0.565 \\
\hline
Low Dosage (\textless{}0.05 mcg/kg/min) & MAE & 8.146 & 8.012 & 9.740 & 9.820 & 7.563 & 7.477 & 8.152 & 8.137 \\
 & AUC & 0.519 & 0.581 & 0.500 & 0.578 & 0.500 & 0.559 & 0.500 & 0.553 \\
\hline
High Titration (\textgreater{}0.05 mcg/kg/min) & MAE & 12.609 & 12.084 & 6.409 & 5.111 & 4.301 & 4.599 & 5.438 & 5.208 \\
 & AUC & 0.350 & 0.650 & 0.500 & 0.550 & 0.500 & 0.583 & 0.500 & 0.708 \\
\hline
Low Titration (\textless{}0.05 mcg/kg/min) & MAE & 6.919 & 6.771 & 9.552 & 9.838 & 7.978 & 7.707 & 8.179 & 8.083 \\
 & AUC & 0.557 & 0.584 & 0.500 & 0.555 & 0.500 & 0.577 & 0.500 & 0.548 \\
\hline
Dose Increase & MAE & 6.982 & 6.58 & 8.265 & 9.083 & 6.673 & 6.760 & 6.654 & 6.742 \\
 & AUC & 0.500 & 0.613 & 0.500 & 0.592 & 0.500 & 0.542 & 0.500 & 0.530 \\
\hline
Dose Decrease & MAE & 7.357 & 7.325 & 10.031 & 9.876 & 8.423 & 7.993 & 8.83 & 8.616 \\
 & AUC & 0.512 & 0.576 & 0.500 & 0.527 & 0.500 & 0.591 & 0.500 & 0.570 \\
 \hline
\end{tabular}
\caption{Comparison of MAE and AUC between the LR and RNN across norepinephrine titrations characterized as an independent application and independent observation, partitioned by titration types.
}
\label{table7}
\end{table}

\begin{table}[ht]
\centering
\begin{tabular}{l  l | rr | rr | rr | rr}
\hline
\multicolumn{2}{c|} {MAE} & \multicolumn{2}{c|} {HR (bpm)} & \multicolumn{2}{c|}{SBP (mmHg)}  & \multicolumn{2}{c|}{DBP (mmHg)}  & \multicolumn{2}{c} {MAP (mmHg)}  \\
&  & LR & RNN & LR & RNN & LR & RNN & LR & RNN \\
 \hline
Dopamine & Survivors & 9.38 & 9.26 & 10.54 & 10.19 & 8.44 & 8.20 & 9.04 & 8.81 \\
 & Non-Survivors & 9.17 & 8.83 & 11.91 & 11.83 & 10.30 & 10.39 & 10.81 & 10.93 \\
 \hline
Epinephrine & Survivors & 8.05 & 7.89 & 12.53 & 12.17 & 9.39 & 9.26 & 10.31 & 10.09 \\
 & Non-Survivors & 7.17 & 7.26 & 9.00 & 9.63 & 8.06 & 8.52 & 9.20 & 9.77 \\
 \hline
Norepinephrine & Survivors & 9.17 & 8.95 & 10.71 & 11.52 & 8.18 & 8.31 & 8.83 & 8.83 \\
 & Non-Survivors & 3.54 & 3.56 & 7.28 & 8.04 & 6.74 & 6.61 & 6.88 & 7.11\\
 \hline
\end{tabular}
\caption{Comparison of LR and RNN MAEs, computed over only those titrations characterized as independent application and an independent observation, partitioned by mortality outcome.}
\label{table8}
\end{table}